\documentstyle[floats,prc,aps,epsf]{revtex}
\begin{document}
\draft
\twocolumn[\hsize\textwidth\columnwidth\hsize\csname@twocolumnfalse%
\endcsname

\title{Sensitivities of the Proton-Nucleus Elastic Scattering Observables 
of $^6$He and $^8$He at Intermediate Energies}
\author{S.P. Weppner and Ofir Garcia}
\address{Natural Sciences, Eckerd College, St. Petersburg, Florida 33711}
\author{Ch. Elster}
\address{Institute of Nuclear and Particle Physics, Ohio University, 
Athens, Ohio 45701}
\date{January 17, 2000}
\maketitle
\begin{abstract}
We investigate the use of proton-nucleus elastic scattering experiments 
using secondary beams of $^6$He and
$^8$He to determine the physical structure of 
these nuclei. The sensitivity 
of these experiments to nuclear structure is examined by using
four different nuclear structure models with different spatial features
using a full-folding optical potential model. 
The results show that elastic scattering at intermediate energies
($<$ 100 MeV per nucleon)
is not a good constraint to be used to determine 
features of structure. Therefore researchers should look elsewhere to 
put constraints on the ground state wave function of the
$^6$He and $^8$He nuclei.
\end{abstract}
\draft

]

\section{INTRODUCTION}
The advent of radioactive accelerator beams 
during the past decade has enhanced
the variety of nuclear reactions available for study. 
In the present research we concentrate on the neutron rich 
isotopes of helium, $^6$He and $^8$He, which
have been produced as secondary beams at intermediate 
energies~\cite{datahe6,korsh}. 
The simplified 
shell model structure of these isotopes 
is thought to be
a core $^4$He, surrounded by loosely bound valence neutrons
located in the $p$ shell, denoted often as the nuclear halo.
A significant amount of research has been done 
on constructing models that reproduce data from 
experimental reactions involving
these
isotopes~(Refs.~\cite{kuo,tostevin2,chulkov,gareev1,gareev2,goncharov},
\cite{gil,bertulani,korsh1,korsh2,karataglidis,garrido2,garrido3,garrido4}
\cite{wurzer1,hiyama,varga,zhukov1,zhukov2,corbis} for example). 
To use radioactive beams effectively  in nuclear studies
the present uncertainty
in the ground state wave functions of $^6$He and $^8$He must be reduced. 
Once the wave functions are known with better precision,
the radioactive beam 
experiments may produce significant 
implications for neutron stars, shell model 
calculations, the two body nuclear force, and the 
three body nuclear force.

One way to ascertain the physical structure of the exotic helium nuclei 
would be to use elastic scattering 
data. 
We want to address the feasibility of this method
by developing proton-$^6$He and proton-$^8$He first order optical potentials 
at intermediate energies (60 MeV-100 MeV per 
nucleon) using different structure models as
inputs to this optical potential model. 
A fair amount of earlier work examines this 
sensitivity~\cite{tostevin2,chulkov,gareev1,gareev2},
\cite{goncharov,gil,bertulani,korsh1,korsh2,karataglidis,kaki}, 
but there is not full agreement in the literature on the strength of this 
sensitivity of structure to elastic scattering data.
For example Ref.~\cite{chulkov}
found that at these energies 
proton-nucleus elastic scattering data was not an 
effectual tool in determining structure. 
Korsheninnikov {\it et al.}~\cite{korsh1,korsh2} also 
did a detailed study on the
sensitivities of proton elastic scattering 
not only of the helium isotopes but also the lithium isotopes 
($^9$Li and $^{11}$Li), using an eikonal approach. They
concluded that elastic scattering ``is not a very promising tool''
to determine structure of the valence neutrons. 
Their belief is that the size of the core
plays a more important role in determining the differential cross section
than the lower density valance neutrons. 
More  recently,  and in contrast,
Karataglidas {\it et al.}~\cite{karataglidis}
have performed calculations on the same exotic helium
reactions at intermediate energy range using a
few different variations of a structure calculation in a g-matrix elastic
optical potential calculation. They concluded for $^6$He
that the data available 
was insufficient for the elastic scattering calculations to discern
the existence of a halo. 
For $^8$He, 
they ascertained
that there was enough data to conclude that it is not a halo nucleus from
comparing differences in the elastic differential cross section calculations.

In the literature
there is not general agreement
to the question of sensitivity of the elastic proton-nucleus 
differential cross section at intermediate energies
to the structure calculation of the target nuclei $^6$He
and $^8$He.  Most research concludes that the 
sensitivity  is not there to determine
structure, but some authors have used elastic proton-nucleus scattering to 
put constraints on the details of their physical 
structure, specifically the halo.
In this work,
we will systematically  examine this 
sensitivity using four independent structure
models and conclude whether
elastic scattering is a tool that should be used to ascertain the
structure of $^6$He and $^8$He.

In section II, we will briefly summarize our full-folding optical potential
calculation technique which we use to describe elastic
proton-$^6$He and proton-$^8$He scattering and we outline the
four different structure models used to describe the helium isotopes.
Our results are in section III and our conclusions are in section IV.

\section{FULL-FOLDING OPTICAL POTENTIAL}

A standard microscopic approach to the elastic scattering of a strongly
interacting projectile from a target of $A$ particles is 
given by the formulation
of an optical potential in `$\tau\rho$' form 
where $\tau$ contains information 
about the nucleon-nucleon interaction and 
$\rho$ is a nuclear structure calculation
(ground state density)
of the target. 
The development of this optical potential begins with
the separation of the Lippmann-Schwinger
equation for the transition amplitude
\begin{equation}
T = V + V G_0(E) T  \label{eq:2.1}
\end{equation}
into two parts, namely an integral equation for $T$:
\begin{equation}
T = U + U G_0(E) P T,  \label{eq:2.2}
\end{equation}
where $U$ is the optical potential operator given through a second
integral equation
\begin{equation}
U = V + V G_0(E) Q U.  \label{eq:2.3}
\end{equation}
In the above equations, the potential operator $V$ represents the external
many-body interaction. The potential operator 
$V={\sum_{i=1}^{A}} v_{0i}$ consists of the
two-body potential $v_{0i}$ acting between the projectile and
the nucleons in the target nucleus. The operators $P$ and $Q$ are projection
operators, $P+Q=1$, and $P$ is defined such that Eq.~(\ref{eq:2.2}) is
solvable. In this case, $P$ is conventionally taken to project onto
the elastic channel. For more details see
Refs.~\cite{density,thesis}.

The evaluation  of the full-folding optical potential
requires a fully off-shell nuclear 
density matrix, which
in its most general form is given as
\begin{equation}
\rho({\bf r'},{\bf r}) = {{\Phi_A}}^\dagger({\bf r'})\;{{\Phi_A}}
({\bf r}) \label{eq:3.1}.
\end{equation}
Here, ${{\Phi_A({\bf r})}}$ is the wave function describing the
nuclear ground state in position space. 
We choose four models for this work to describe the $^6$He and $^8$He 
ground state. The structure 
models vary in rigor, quality, and applicability in 
describing these exotic helium nuclei.
First, a general description of each model will be given in part A,
to be followed by comparisons
of all four models in part B.

\subsection{Descriptions of the four off-shell densities}

Our first model, proposed by Sherr~{\cite{sherr}},
will be referred to as the 
 `boot-strap' model (BS). 
This model was created to describe the root mean
squared (rms) radii of a variety of exotic nuclei using
a simple description of the nucleus. It represents 
the valence neutrons by using a Woods-Saxon potential
which is fit to the two neutron binding energy. It then follows a
sequential step
procedure to build the exotic nuclei. Explicitly for helium, the
model starts with the 
well known core $^4$He, and then 
builds $^6$He by calculating the wave function
for the valence neutrons. Likewise, to construct $^8$He, 
$^6$He is considered the
core and the 2 neutron wave function generated from a Woods-Saxon potential
is calculated.
Wave functions for the $^4$He core and
valence neutrons were calculated in r space (relative to the center of mass of the whole
nucleus),
then Fourier transformed to 
momentum space
\begin{equation}
\rho'({\bf p'},{\bf p}) = \frac{1}{8\pi^3}
\int d^3{\bf r'} e^{-i{\bf r'}\cdot {\bf p'}}
\int d^3{\bf r} e^{-i{\bf r}\cdot {\bf p}} \; \rho({\bf r'},{\bf r})
\label{eq:3.4},
\end{equation}
where they were used in construction of the off-shell density
via Eq.~({\ref{eq:3.1}}).
The validity
of this model is questionable due to its extreme simplicity. 
It is thought that the size of the
$^4$He core in the exotic isotopes is larger than the bare $^4$He radius which
this model assumes~\cite{zhukov1}.  
The mode of construction of $^8$He is also 
in contradiction with most other structure calculations for $^8$He. To
first order, $^6$He could be approximated as a $^4$He + 2n; however, it is
apparent that $^8$He is closer to $^4$He+4n than ($^4$He +2n) +2n as the BS
model would suggest~\cite{zhukov1}. 

The second model is a relativistic point coupling model within the
framework of a chiral effective 
field theory by Rusnak and Furnstahl~\cite{chiral1}.
A Lagrangian is constructed: an expansion in powers of the scalar,vector,
isovector-vector, tensor, and isovector-tensor densities, 
and their derivatives. 
The theory contains all the symmetries of QCD and is able to
calculate low energy features, such as the structure of nuclei 
ground states adequately.
For this paper, we used what Ref.~\cite{chiral1} 
refers to as the `FZ4' scheme. Here,
the vector meson and $\rho$ meson 
masses are fixed, while the coefficients of the
densities to fourth order are varied to 
produce a low chi-square to experimental
observables. 
Most 
varieties of the chiral effective theory reproduce the
bulk nuclear observables
of spherically symmetric nuclei.
Questions of applicability to exotic helium nuclei
can of course be raised while using this model,
for it was not developed for the
non-spherical, non-bulk nuclei $^6$He and $^8$He.
The numerical procedure to 
create the off-shell densities~(Eq.~\ref{eq:3.1}) used for the optical
potential is given in Ref.~\cite{density}. 

The third nuclear structure model used to 
describe the densities $^6$He and $^8$He,
a Dirac-Hartree model (DH)\cite{DH1,DH2},
has been used extensively by two of
the authors (Ch.~Elster and S.~P.~Weppner)
to describe the structure of doubly magic 
spherical nuclei with 
success~\cite{density,thesis,med2,med1,energydep,65mev}.  This is 
the oldest structure model of the four 
models discussed. 
The FZ4 model (detailed above)
has the same structure
wave function, so the method used to create the momentum 
off-shell density of Eq.~(\ref{eq:3.1}) is the same for both models
and is detailed in Refs.~\cite{density,thesis}. Applicability is a concern for
this model also. When developed, it was fit to the bulk properties of 
$^{16}$O, $^{40}$Ca, $^{48}$Ca, $^{90}$Zr, and $^{208}$Pb, all doubly
magic nuclei. Furthermore, this 
model, as well as FZ4, falls under the mean field ansatz,
which is rather tenable when describing
nuclei with only 6 or 8 nucleons. It was also developed well before the 
general structure of $^6$He and $^8$He were apparent, thereby making it
a candidate to test whether the elastic observables can detect
this non-applicability.

\begin{figure}[t]
\makebox{
\setlength\epsfxsize{8.4cm}
\epsfbox{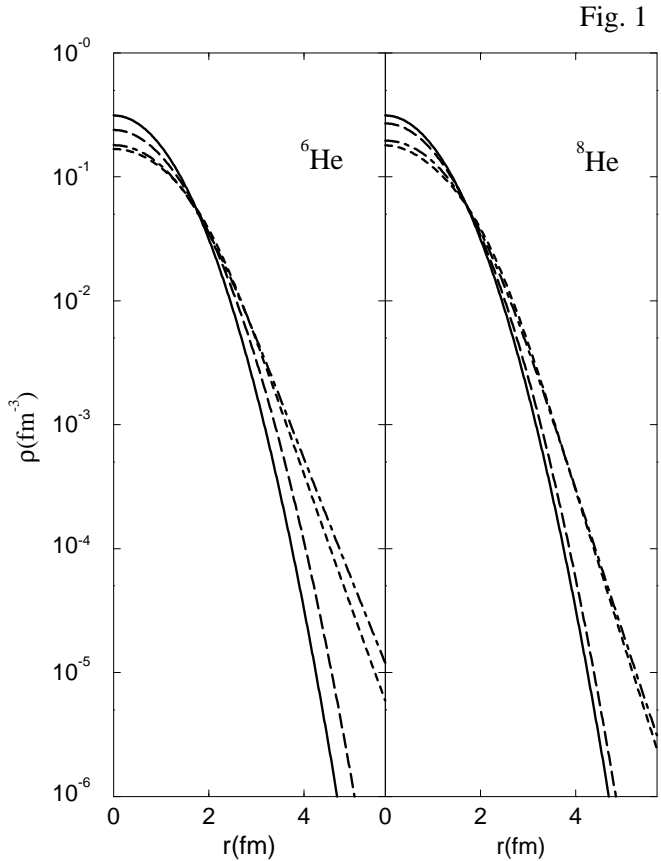}}
\vspace*{-1cm}
\caption{ Proton density calculations in position space for
$^6$He and  $^8$He.
The solid line represents the calculation performed with a
Boot
Strap nuclear structure model~\protect\cite{sherr}.
The short dashed line represents
the same calculation using a
chiral model~\protect\cite{chiral1,chiral2} for the
nuclear structure. The dash-dotted line represents the
same calculation using a Dirac-Hartree
model~\protect\cite{DH1,DH2} as the
nuclear structure calculation and the long dashed line has a
cluster
orbit shell model approximation (COSMA)~\protect\cite{zhukov1,zhukov2}
as a model. \label{fig1}}
\end{figure}

\noindent
\begin{figure}[t]
\makebox{
\setlength\epsfxsize{8.4cm}
\epsfbox{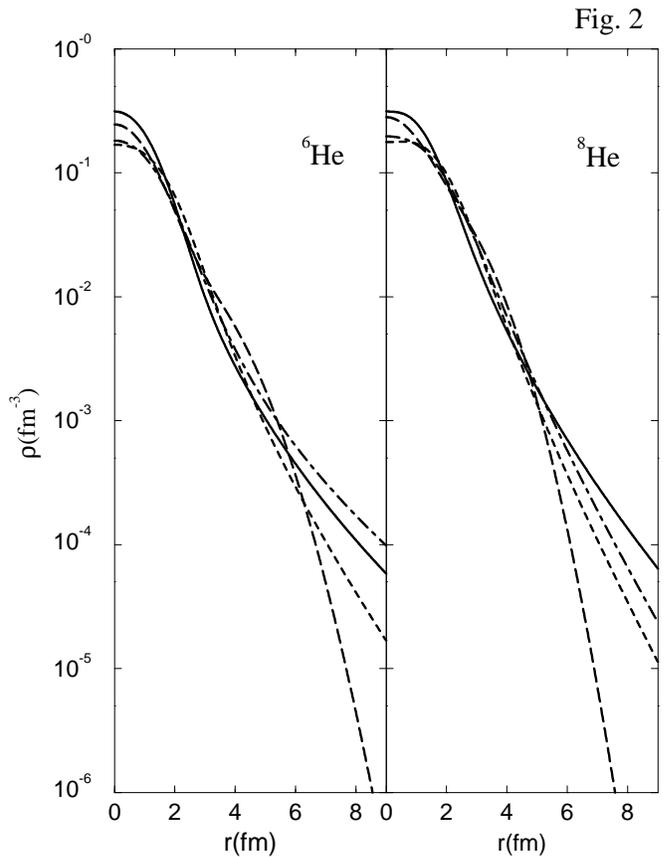}}
\vspace*{-1cm}
\caption{The legend is the
same as Fig.~\protect\ref{fig1} except the lines now represent
calculations of neutron densities.
\label{fig2}}
\end{figure}

The last nuclear model to be discussed was developed explicitly for exotic
nuclei. The COSMA model (cluster 
orbit shell model approximation)~\cite{zhukov1,zhukov2},
an approximation to the three body problem,
has been used extensively in the literature to describe 
elastic scattering with the exotic helium 
nuclei~\cite{tostevin2,gareev1,gareev2,korsh1,korsh2,tostevin1,crespo}.
The COSMA model is a combination of nucleon clustering and the standard
shell model, which obeys the Pauli exclusion principle by using Slater
determinants to 
produce a fully antisymmetrized wave function in r-space. They are then
translated as single particle wave functions relative to the center of mass
of the entire nucleus.
These
wave functions are Fourier transformed into momentum space using
the method of Eq.~(\ref{eq:3.4}), thus making it
possible to create a fully off-shell density 
in momentum space. 

There have been more rigorous models developed 
for $^6$He and $^8$He (Refs.\cite{korsh,wurzer1,hiyama,varga,zhukov1,garrido1}
among others) which we have not used. In all cases, these models
treat the three body problem (core + nucleon + nucleon) or
(core +di-neutron +di-neutron) with fewer approximations than the models
presented here. Our goal is to present a sensitivity test, 
rather than produce the most
fundamental calculations possible. To that end, we have used four highly 
varying models, with different characteristics, which are easy to calculate.
As an aside, in a comparison 
between a realistic three body approach (treating the antisymmetry
correctly) and the reducible
two body approximation, the authors of Ref.~\cite{garrido1} note little
difference at intermediate and long ranges. They do find small differences 
in the short range behavior of the wave functions, but these are of little
consequence when describing 
intermediate energy elastic scattering. Therefore we would expect
our results using these simpler models to differ little from results using
the more rigorous realistic models.

\subsection{Nuclear structure of the four models}

\begin{table*}[t]
\begin{tabular}{|l||cccccc|cc|}
&\multicolumn{6}{c}
{\bf $^6$He}&& \\
\hline
Model& $r_{rms}$ [fm] & $r_c$ [fm]  &$\Delta
 r_c$ [fm] & $r_v$ [fm] & $\Delta r_v$ [fm]
& $S_{2n}$ [MeV]& $S_d$ [fm]& halo? \\
\hline
BS  & 2.90  & 1.61 &0.63 & 4.47  & 2.59 &0.98 &-0.36 & no \\
FZ4  & 2.54 & 1.98 &0.81 &3.39  & 1.52 &1.80 &-0.92  &no\\
DH  & 3.75  & 2.00 &0.84 &5.84  & 3.49 &0.19 &-0.49 & no  \\
COSMA & 2.57& 1.77 &0.69 &3.68  & 1.20 &--& +0.02& yes \\
EXP  &  2.39&  -- & -- &  -- &  -- & 0.97 &? &? \\
\hline
&\multicolumn{6}{c}{\bf $^8$He} &&\\
\hline
Model& $r_{rms}$ [fm]  & $r_c$ [fm] &$\Delta
 r_c$ [fm] & $r_v$ [fm] & $\Delta r_
v$ [fm] & $S_{4n}$ [MeV]& $S_d$[fm] &halo? \\
\hline
BS & 2.84 & 1.61 & 0.63 &3.68 & 1.91&3.12&-0.47 & no \\
FZ4& 2.57 & 1.95 & 0.80 &3.08 & 1.24&3.47&-0.91 &no  \\
DH & 2.75 & 1.90 & 0.79 &3.39 & 1.41&2.68 &-0.71 & no\\
COSMA &2.52&1.69 & 0.66 &3.14 & 0.99 &--& -0.20 & no  \\
EXP   &2.49& --    & --     & --    & --     &3.1  &? & ?
\end{tabular}
\caption{Comparison of the four structure
models observables with each other and
experiment.
The root mean squared matter (rms) radius ($r_{rms}$) of the whole
nucleus, the rms radius of the
two neutron-two proton core wave function($r_c$),
the standard deviation of the core ($\Delta r_c$),
the valence neutron matter radius ($r_v$),
the standard deviation of the valence
neutron wave function ($\Delta r_v$), the
separation energy of the valence neutrons ($S_{2n}$), the separation distance
of one standard deviation
of the core and valence wave functions ($S_d$),
and a statement on whether the model has
a halo structure as defined by this work. The experimental results are from
Ref.~\protect\cite{tanihata}.}
\label{T1}
\end{table*}

In Fig.~\ref{fig1}, we have plotted the proton densities 
in r-space of all four structure models described above.
The general characterization
is that the core two protons and two neutrons are more tightly bound in the 
two models that are designed for exotic nuclei, these being 
the BS model (solid line)
and COSMA model (long dashed line). 
These models have a core close to that of 
the lone $^4$He nucleus ($\approx 1.6$ fm). 
In contrast, the FZ4 model (short dashed line) and
the DH model (dashed-dotted line) have a core which ranges from
12$\%$ to 20$\%$ larger than their exotic counterparts.

The total neutron densities in r-space have also been plotted for
all four structure models for $^6$He and $^8$He in Fig.~\ref{fig2}.
Comparing the exotic nuclei structure models first,
the BS model (solid line) and
COSMA model (long dashed line) have a tight two-neutron core because their
densities are higher in the 0 fm to 2 fm range. At about 3 fm, differences
begin to emerge between these two exotic models.
All models do have an 
extended neutron wave function of varying 
degrees, with the COSMA model having 
the most  unique shape.

In Table~{\ref{T1}}, we list the four models and the characteristics they
describe. All position measurements are
relative to the center of mass of the $^6$He or $^8$He system.
Calculated in Table~{\ref{T1}} are the root mean 
squared matter radius ($r_{rms}$) of the whole
nucleus, the rms radius of the 
two neutron-two proton core wave function($r_c$),
the standard deviation of the core ($\Delta r_c$),
the valence neutron matter radius ($r_v$),  
the standard deviation of the valence
neutron wave function ($\Delta r_v$), the
separation energy of the valence neutrons, the separation distance
of one standard deviation
of the core and valence wave functions, 
and a statement on whether the model has
a halo structure as defined by this work. 
The definition of the standard deviation
is 
\begin{equation}
\Delta r = \sqrt{<r^2>-<r>^2},
\end{equation}
whereas the separation distance is defined as
\begin{equation}
S_d= r_v-\Delta r_v - \Delta r_c - r_c.
\end{equation}
Simply if $S_d >0$ then we define this nucleus as having a halo
because there is a well defined 
separation between the core and halo centers.
The
only discernible halo nucleus is the COSMA model of $^6$He. 
It contains a tightly bounded
core wave function with adequate spacing between core and valence nucleons.
The COSMA model also has a significantly 
different asymptotic
wave function shape. The other three models do not define a halo
for the $^6$He nucleus, as there is too much significant overlap between
the core and valence 
wave functions. 
No model produces a definitive halo for
$^8$He, although COSMA comes closest. 

In summary it is concluded that the 
most disparate structure is the COSMA model. 
One would expect
that because this model is used often to describe exotic nuclei, it
should also best describe elastic scattering if the observables are sensitive
to the nuclear 
structure calculation.  
In the next section we will use these
four models as input into our optical 
potential to describe elastic scattering 
at the intermediate energy range.

\section{RESULTS AND DISCUSSION}
The elastic scattering of protons
from $^6$He and $^8$He at incident proton energies from 66 MeV to
100 MeV are calculated.
At this energy range, 
the proton-nucleus elastic scattering 
data are scarce and exist only at forward angles. 
We will focus on the reactions where 
experimental data for the elastic differential cross-sections exists, 
but we will
also comment on  how our conclusions would change if the experimental
database were enlarged.
Other observables which are calculated in this work,
for which no data exists, are the spin
rotation function ($Q$) and the analyzing power ($A_y$). We will also comment
on 
sensitivity to these observables.

\begin{figure}[t]
\makebox{
\setlength\epsfxsize{9.2cm}
\epsfbox{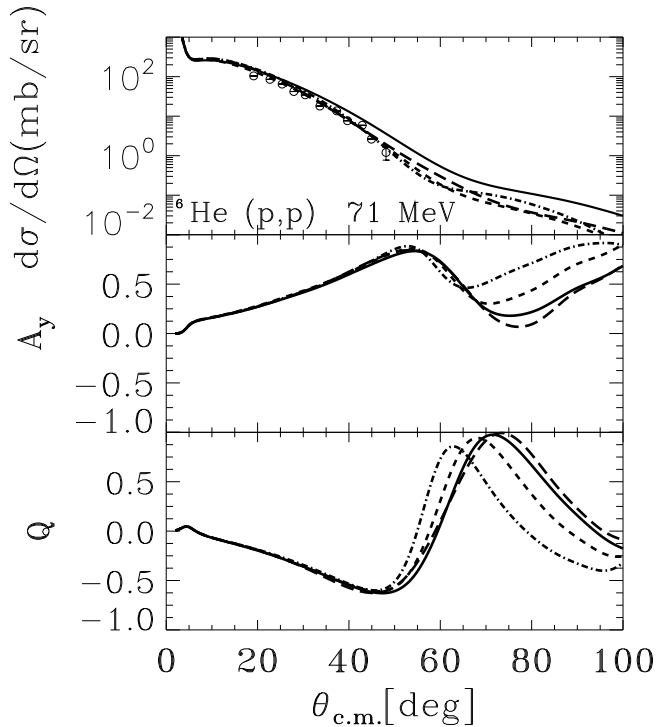}}
\vspace*{.3cm}
\caption{The angular distribution of the differential cross-section
         ($\frac{d\sigma}{d\Omega }$), analyzing power ($A_y$) and
         spin rotation function ($Q$) are shown for elastic proton
         scattering from $^{6}$He at 71 MeV laboratory energy.
All calculations are denoted with the same legend as
Fig.~\ref{fig1}.
The solid line represents the calculation performed with a
first-order full-folding optical potential using the Boot
Strap nuclear structure model~\protect\cite{sherr} as an
input to the optical potential.
The short dashed line represents
the same calculation using a
chiral model~\protect\cite{chiral1,chiral2} for the
nuclear structure of $^6$He. The dash-dotted line represents the
same calculation using a Dirac-Hartree
model~\protect\cite{DH1,DH2} as the
nuclear structure calculation and the long dashed line has a
cluster
orbit shell model approximation (COSMA)~\protect\cite{zhukov1,zhukov2}
as a model. All calculations use Nijmegen I for their
NN interaction~\protect\cite{nijmegen}.
The data (circles) are taken from
Ref.~\protect\cite{datahe6}. \label{fig5}}
\end{figure}

The 
full-folding optical potential used for these results
is calculated as outlined in 
Refs.~\cite{density,thesis}, and we use the four model densities
as described in Sec. II. We will refer to the model of 
Ref.{~\cite{sherr}} as `BS'. The Dirac-Hartree  model of  
Refs.~\cite{DH1,DH2} will be labeled `DH'. The chiral 
point coupling model of 
Ref.~\cite{chiral1,chiral2} will be labeled `FZ4'. The
cluster model of Refs.~\cite{zhukov1,zhukov2} will be referred to as
`COSMA'.

\begin{figure}[t]
\makebox{
\setlength\epsfxsize{9.2cm}
\epsfbox{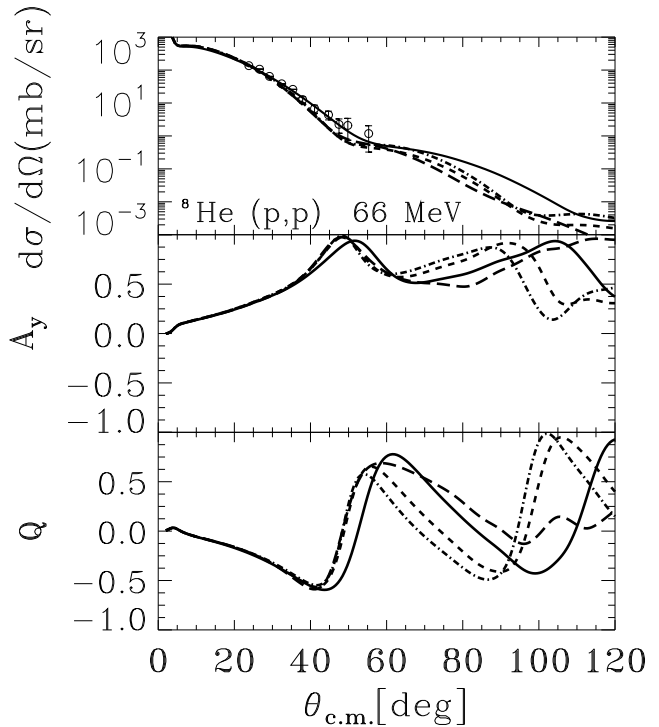}}
\vspace*{.3cm}
\caption{Same as Fig.~\protect\ref{fig5}, except that the reaction is a
proton at 66 MeV laboratory energy elastically scattering
from $^{8}$He.
The data are taken from Ref.~\protect\cite{datahe6}. \label{fig6}}
\end{figure}

The full-folding optical potential also requires a model of the 
NN interaction. In this work, we use the Nijmegen I interaction
\cite{nijmegen}.
We have also calculated some optical
potentials using the CD Bonn NN potential~\cite{cdbonn}. This
potential has the same tight constraints of the Nijmegen 
for its on shell
values to agree with np and pp data, but the off shell amplitudes 
are different.  The elastic scattering calculations
using the CD Bonn potential
show very little difference with those that use the Nijmegen potential. 
The conclusions drawn in this work
are therefore independent of the choice of which modern NN potential was used.

\subsection{Elastic scattering results: effects of structure}

The scattering observables
for elastic proton scattering from $^6$He at 71 MeV are 
displayed in Fig.~\ref{fig5}.
There are four calculations on the figure (using the same legend as 
Figs. \ref{fig1}-\ref{fig2}). 
The solid line represents the elastic differential cross-section
calculated from a full-folding optical potential
using the BS model as the structure calculation, and the short-dashed line
represents a calculation of the observables from a
full-folding optical potential using the FZ4 model as the structure
calculation. The DH version of the 
calculation is represented by the dot-dashed
line while the optical potential using the COSMA structure model was
used in the calculation of  
the long-dashed line. All models use the
Nijmegen I nucleon-nucleon interaction. 
The experimental data for this reaction
were given in Ref.~{\cite{datahe6}}. As also seen in Fig.~\ref{fig1},
the only calculation which does not 
adequately describe the 
limited experimental data for this reaction is  the BS model.  
Referring to Table~\ref{T1}, the only
significant difference between this model 
and other models is the extremely small
core which mimics  the size of a lone $^4$He nucleus.
The other three models agree favorably 
with the limited data, yet when one looks at their features there are
large differences in their 
binding energies, 
rms radii, and presence of a discernible halo. 
Therefore, it is impossible
to draw any conclusions about the structure of $^6$He valance neutrons
from this reaction. 
We may draw some inferences on the appropriate
size of the core from looking at the results produced using the BS model, but 
the shape and existence of a halo cannot be determined from this reaction.

In Fig.~\ref{fig6}, we calculate elastic scattering 
from $^8$He at a projectile energy
of 66 MeV. The legend represents the same calculations 
as in Fig.~\ref{fig5}. The experimental
data are from Ref.~{\cite {datahe6}}. All four models adequately 
represent the data. The
one model with the most significant differences 
in shape is again the BS model (and only at higher angles). This
model has the tightest core, and a 
loose valence 
wave function, as it had for $^6$He. 
There is nothing that can be learned about the 
valance structure of $^8$He from 
studying this reaction's differential cross section for the 
data which exist. 
Polarization measurements (specifically
$Q$) may be used if polarized experiments are done at large angles to high
accuracy ($>$60$^o$).

\begin{figure}[t]
\makebox{
\setlength\epsfxsize{9.2cm}
\epsfbox{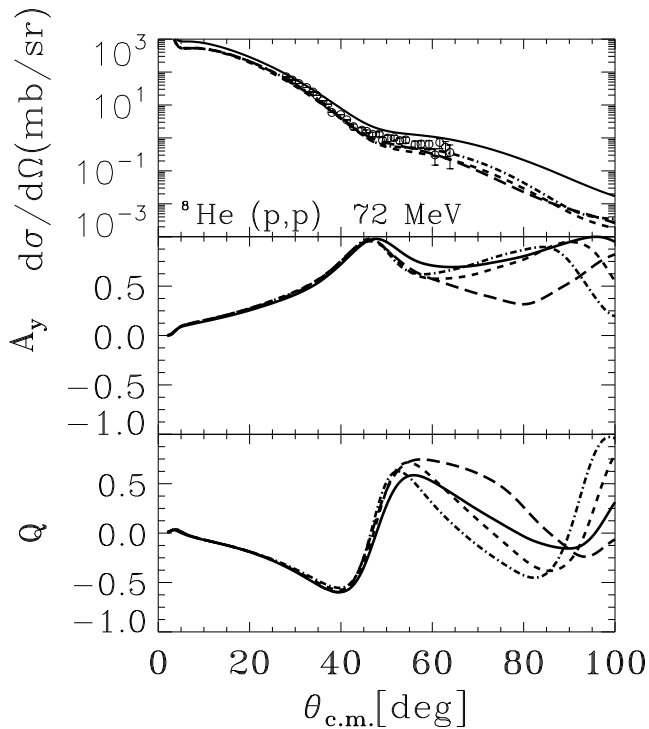}}
\vspace*{.3cm}
\caption{Same as Fig.~\protect\ref{fig5},
except that the reaction is a
proton at 72.5 MeV laboratory energy elastically scattering
from $^{8}$He.
The data are taken from Ref.~\protect\cite{korsh}. \label{fig7}}
\end{figure}

We move to a slightly higher projectile energy in Fig.~\ref{fig7}, 
where protons are scattered
from $^8$He at a projectile energy of 72.5 MeV. 
Again, the calculations have
the same legend as given in Figs.~\ref{fig5} and
Fig.~\ref{fig6}. The data for
this calculation are the most extensive in this energy range, they
approach
the 65$^o$ center of mass angle. 
Unfortunately the structure of $^8$He still cannot be 
determined from this experiment.
The sensitivities due to the structure calculation are not 
strong enough, given the
experimental error, to ascertain the structure of $^8$He. According to
Table~\ref{T1},
the COSMA model has a more defined valence ring than the others, 
but the differential
cross section experimental data are unable to differentiate the validity of
any of these disparate models unless experimenters were
able to measure the differential cross-section with a margin of error of 
less than 20\% at angles above 65$^o$.

\begin{figure}[t]
\makebox{
\setlength\epsfxsize{9.2cm}
\epsfbox{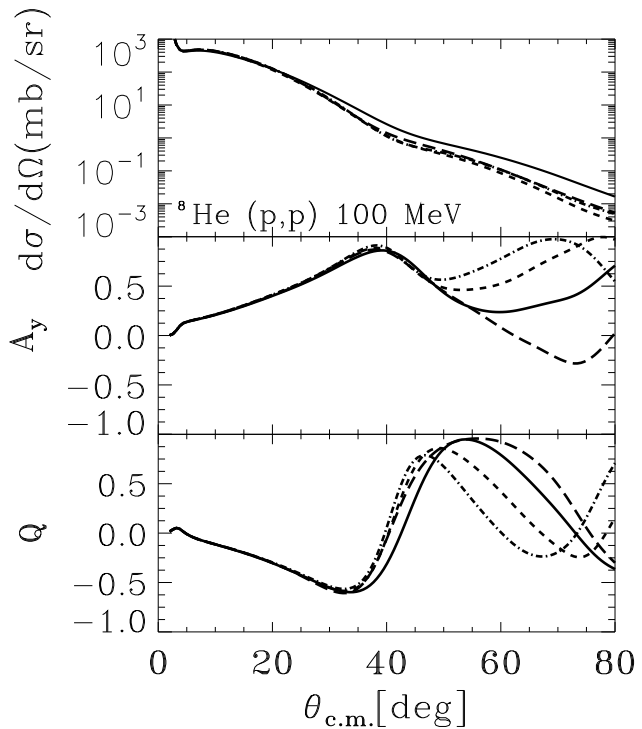}}
\vspace*{.3cm}
\caption{Same as Fig.~\protect\ref{fig5},
except that the reaction is a
proton at 100 MeV laboratory energy elastically scattering
from $^{8}$He.
No data exists for this reaction. \label{fig8}}
\end{figure}

In Fig.~\ref{fig8}, we explore a higher energy reaction where no 
experimental data exist.
Here, we calculated protons scattering  from $^8$He at 100 MeV. Once more,
the calculations use the same models and line definitions as 
in Figs.~\ref{fig5}-\ref{fig7}.  Again, the 
three structure calculations give 
similar results (DH,FZ4, COSMA). The BS model
runs high through the whole calculation, similar in significance to the
previous figures. 
Polarization experiments 
also tell us little below 70$^o$ scattering 
center of mass angle. Even at 100 MeV the elastic reaction is insensitive to
the structure of the valence neutrons in $^8$He.

Recently, in Ref.~\cite{karataglidis}, a similar study was done using these
elastic reactions and pion production. 
The authors concluded that neither 
$^6$He nor $^8$He were halo nuclei. 
Their inclusion of a halo in their structure
calculations  always lowered the differential
cross section at angles greater than 15$^o$.  In this present work, we did not
find such a simple relationship between non-halo and halo nuclei. In fact, 
the most extreme halo
model of $^6$He (dashed line - COSMA) was not at either extreme of
the differential cross section calculation of Fig.~5. 
We therefore conclude that the differential cross section
is only slightly sensitive to the existence of a halo ($S_d$),  
the spread of 
the halo ($\Delta r_v$), the radius of the core nucleons ($r_c$), 
and the binding energy of the core nucleons ($\Delta r_c$), all of which are
coupled to each other in a complex fashion. 
The structure parameter 
that seems to have the most influence at this energy
is the radius of the core $^4$He particle. 
There seems to be almost complete 
insensitivity to the valence neutron wave
functions. To reiterate, the COSMA model has a very distinct asymptotic shape 
for the valence neutrons, and this uniqueness does not transfer
into the differential cross section as exhibited by the similarity in 
the calculations.

\subsection{Medium effects}

So far in this work, we have used the impulse approximation,
setting the medium field to zero. 
In previous work two of the authors (Ch. Elster and S.~P.~Weppner)
showed that at 65 MeV, if a medium field was used (as outlined  
briefly in Section II, 
and in more detail in Refs.~\cite{med2,med1,65mev}) then 
there was a systematically better fit with elastic scattering observables
across a wide range of stable spin-0 nuclei.  

For two of the structure
calculations, the DH and the FZ4, we added
a mean field consistently. If we used a DH structure model then
we used the same DH model to simulate our mean field; likewise, this 
was also done using the FZ4 structure model. 
Overall the effects of adding this mean field to the  $^6$He and $^8$He
calculations of elastic scattering observables were 
smaller than seen
previously for other nuclei.

\begin{figure}[t]
\makebox{
\setlength\epsfxsize{9.2cm}
\epsfbox{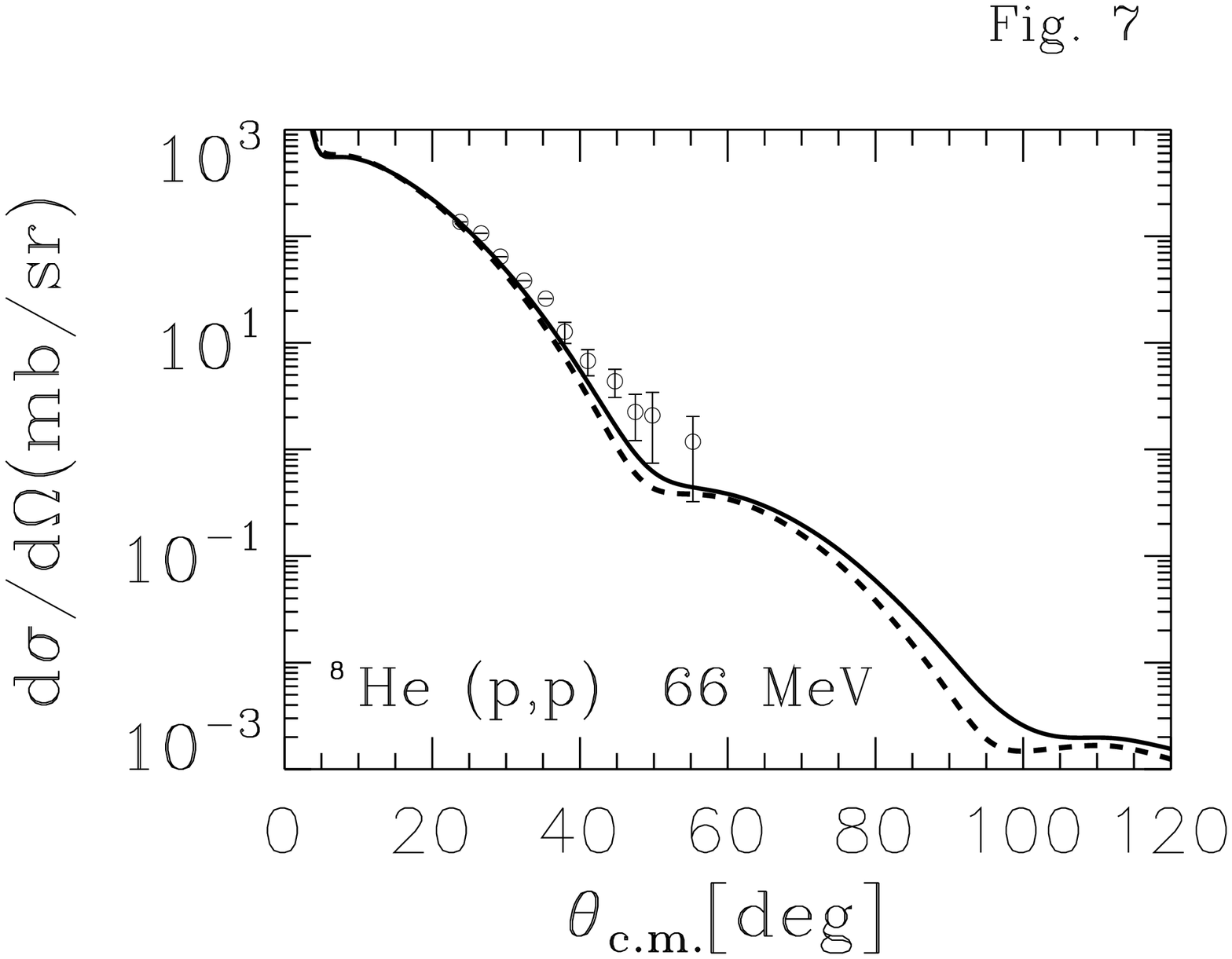}}
\vspace*{-3.25cm}
\caption{ The angular distribution of the differential cross-section
         ($\frac{d\sigma}{d\Omega }$)
         is shown for elastic proton
         scattering from $^{8}$He at 66 MeV laboratory energy.
The solid line is a calculation without medium effects, the short
dashed line has medium effects included.
Both calculations use a chiral model~\protect\cite{chiral1,chiral2}
for the nuclear structure calculation of $^8$He and use the
Nijmegen I potential~\protect\cite{nijmegen} as their NN interaction.
The data (circles)
are taken from Ref.~\protect\cite{datahe6}. \label{fig14}}
\end{figure}

\begin{figure}[t]
\makebox{
\setlength\epsfxsize{9.2cm}
\epsfbox{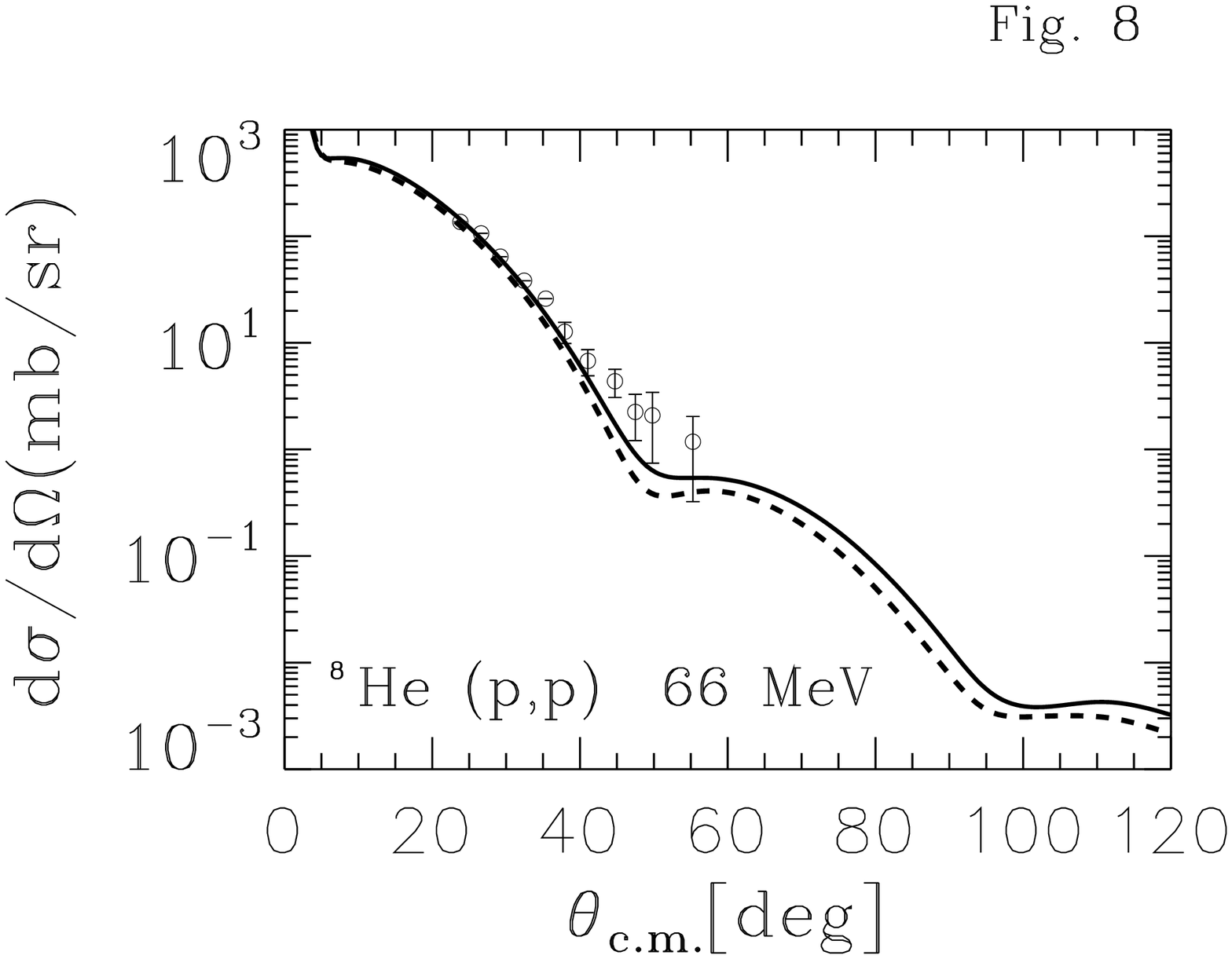}}
\vspace*{-3.25cm}
\caption{Same as Fig.~\protect\ref{fig14},
except both calculations use a Dirac Hartree
calculation~\protect\cite{DH1,DH2} to
model $^8$He. \label{fig15}}
\end{figure}

In Fig.~\ref{fig14}, 
we compare two calculations of $^8$He elastically scattering
off a proton at 66 MeV. Both calculations use the DH structure calculation
and the Nijmegen I interaction. The difference is that the solid
line sets the mean field to zero, while the dashed line includes it.
For comparison, in Fig.~\ref{fig15},
the same calculation is done using the FZ4 
structure calculation and mean field using the same FZ4 model. 
Both calculations give the same results:
when the medium effect is added, it 
systematically lowers the differential cross section slightly. In general, the
effect is smaller than for larger spin-0 nuclei previously  
studied~\cite{med2,med1,65mev}. Since
these are smaller nuclei, and less tightly bound, this conclusion seems
reasonable. However, it is important to note that this
small change did not lead to a better description of the experimental data,
in contrast to earlier work with other nuclei 
where there was a systematic improvement.

\begin{figure}[t]
\vspace*{1cm}
\makebox{
\setlength\epsfxsize{9.2cm}
\epsfbox{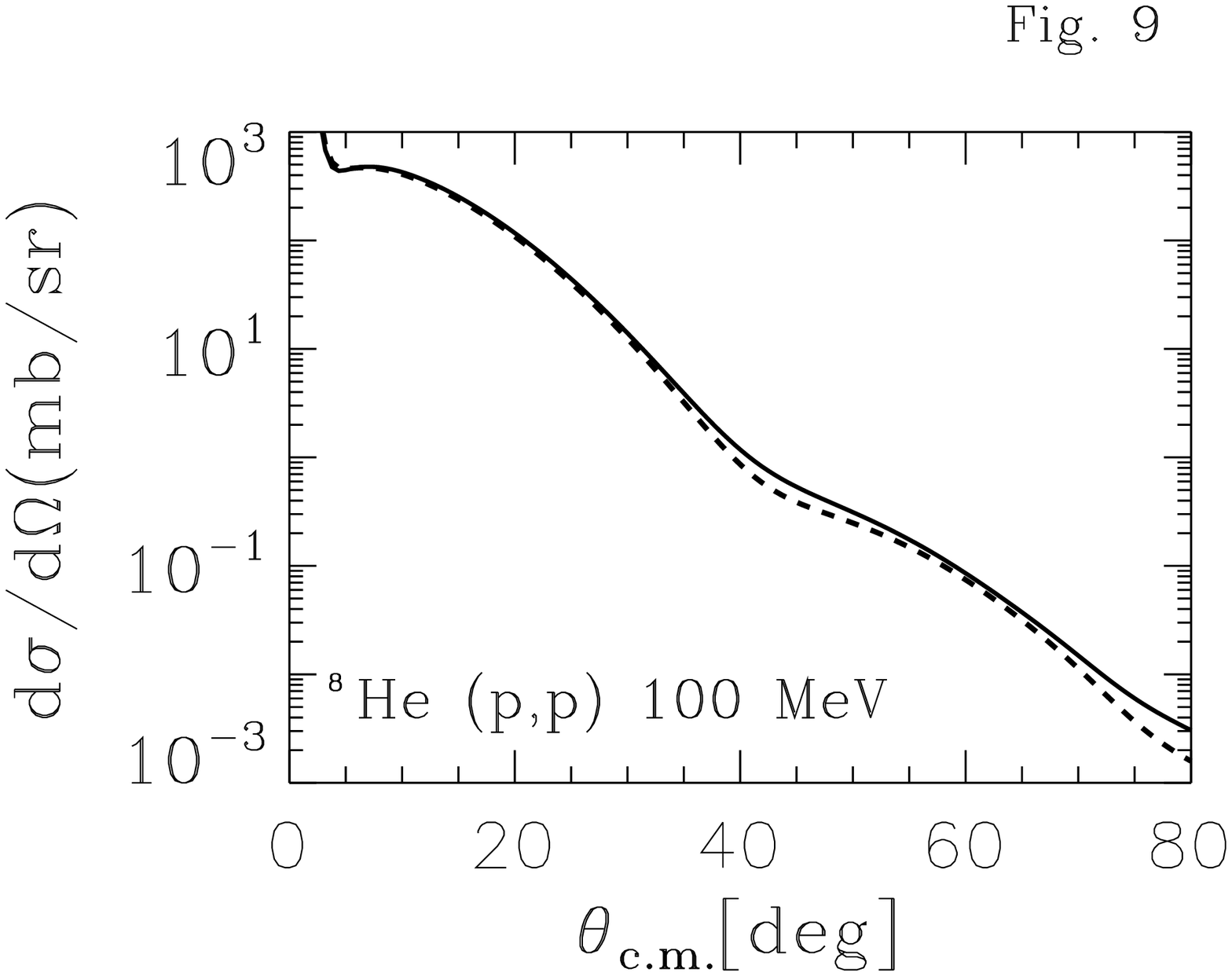}}
\vspace*{-3.25cm}
\caption{Same as Fig.~\protect\ref{fig14},
except that now the reaction is at 100 MeV laboratory energy.
There is no data for this reaction. \label{fig16}}
\end{figure}

\begin{figure}[t]
\makebox{
\setlength\epsfxsize{9.2cm}
\epsfbox{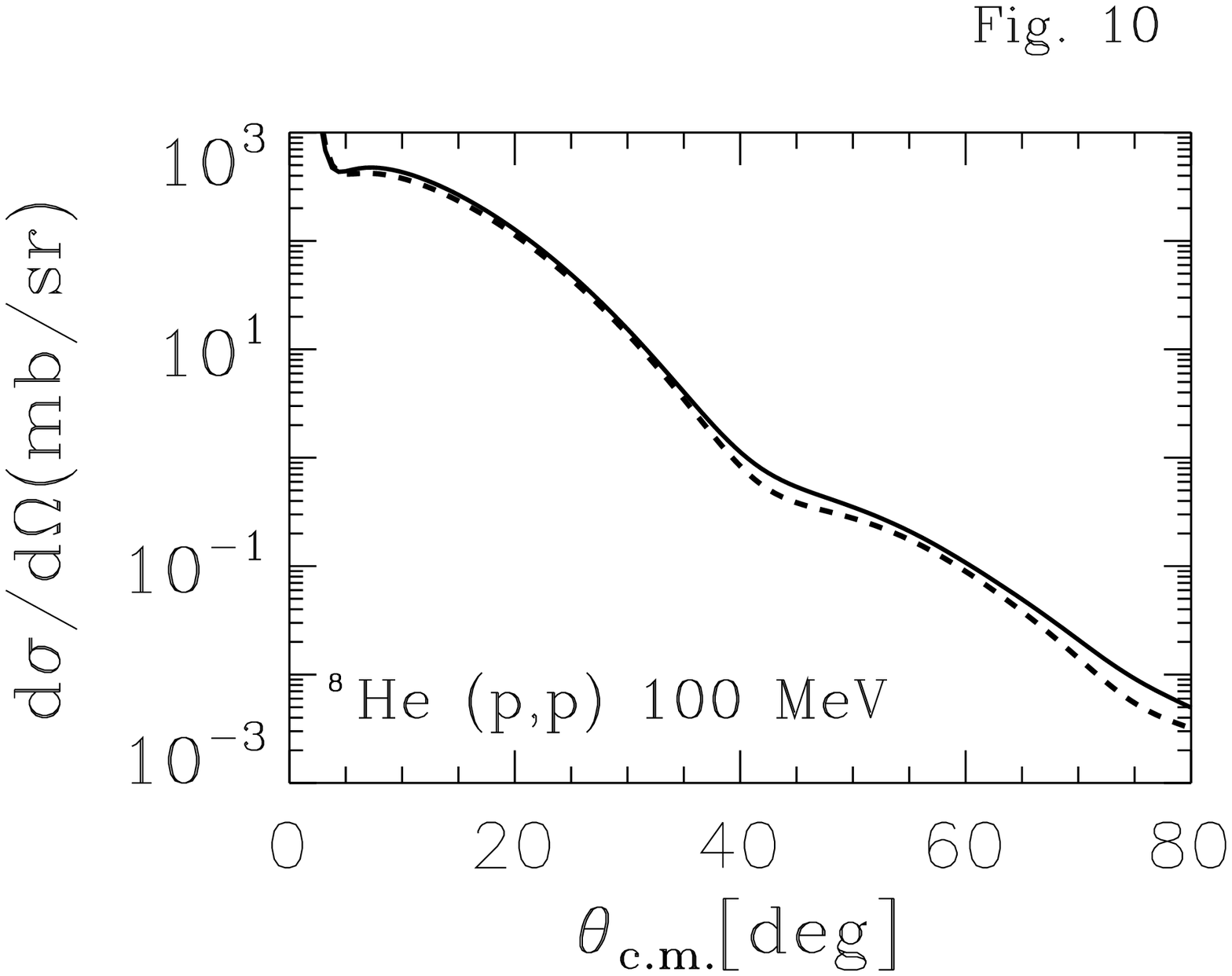}}
\vspace*{-3.25cm}
\caption{Same as Fig.~\protect\ref{fig15},
except that now the reaction is at 100 MeV laboratory energy.
There is no data for this reaction. \label{fig17}}
\end{figure}

At higher energies, these trends continue, although their effects are smaller.
We plot in Fig.~\ref{fig16} the elastic observables
of $^8$He at an energy of 100 MeV colliding 
with a proton. As in Fig.~\ref{fig14},
the solid line represents the DH calculation without mean field effects,
while the dashed line includes the effects. Both calculations use the 
Nijmegen I potential. These medium effects are barely discernible at this
higher energy. This trend has been seen before in earlier work with
other nuclei~\cite{med2,med1}. For completeness, in  
Fig.~\ref{fig17} we have calculated 
the same reaction as Fig.~\ref{fig16} except we now
use the FZ4 structure calculation and mean field (dashed line)
for $^8$He. The same conclusions are 
reached. By using two different models, we 
conclude that this mean field procedure leads to results that are model
independent and smaller than doubly magic nuclei at the same energies.

\section{SUMMARY and CONCLUSION}

We have presented sensitivity tests for elastic scattering 
observables of protons bombarded with
$^6$He and $^8$He. 
Here, we found that elastic scattering is a weak tool for determining
the structure of these isotopes. These conclusions were drawn by using
four different nuclear structure models that had different spatial 
characteristics in the calculation of the proton-nucleus optical potential.
All calculations using these structure models 
were in good agreement with the data that exist. In fact,
the models not designed for exotic nuclei (Dirac-Hartree and chiral models) 
did as well as, and
sometimes slightly better than their made-for-exotic-nuclei counter parts
(COSMA and a simple `boot-strap' model).
We agree with the results of earlier work of 
Korsheninnikov {\it et al.}~\cite{korsh1,korsh2}.
They believe that  the size of the core
plays a more important role in determining the differential cross section
than the lower density valance neutrons.
The only potential area for significant nuclear structure sensitivity with
elastic scattering is with the large angle ($>$ 70$^o$) spin observables.
Since the radioactive beams are secondary beams, to  produce enough
polarized statistics  to measure these reactions with any
accuracy is beyond experimental
and theoretical capabilities at the present time. 
It is, therefore, possible to conclude that one should look
beyond intermediate elastic reactions when trying to determine the 
structure of the neutron
rich helium  isotopes.
Higher energy elastic scattering ($>$ 500 MeV/nucleon)~\cite{tostevin2} 
has had some success in
determining structure, although 
they warn against using an optical model approach, as used 
here~\cite{crespo}. Inelastic hadron reactions~\cite{thompson}
(momentum distributions following
fragmentation~\cite{garrido4,tanihata,zhukov3}, 
transfer reactions~\cite{transfer1}, 
Coulomb breakup~\cite{coulumb1,coulumb2}, excitation~\cite{danilin}, 
and charged 
pion photo production~\cite{pion1})
and
an interesting concept using electron scattering~\cite{garrido2}
offer hope as tools
to determine conclusively the structure of $^6$He and $^8$He.

\acknowledgments

This work was performed in part under the auspices of the National
Science Foundation under grant No. PHY-9804307 with
Eckerd College, the auspices of Eckerd College 
under the Hughes Foundation, and the 
auspices of the Department of Energy 
under grant No. DE-FG02-93ER40756 
with Ohio University.
We thank the National Partnership for Advanced 
Computational Infrastructure (NPACI) under grant No.~ECK200 and the 
Ohio Supercomputer Center (OSC) under grant No.~PHS206  for
the use of their facilities.  We would also like to 
thank Richard Furnstahl for the use of his 
chiral structure code.


\begin{thebibliography}{10}

\bibitem{datahe6}
A.~A. Korsheninnikov {\it et~al.}, Phys. Rev. C {\bf 53},  R537  (1996).

\bibitem{korsh}
A.~A. Korsheninnikov {\it et~al.}, Phys. Lett B. {\bf 316},  38  (1993).

\bibitem{kuo}
T.~T.~S. Kuo, F. Krmpoti\'{c}, and Y. Tzeng, Phys. Rev. Lett. {\bf 78},  2708
  (1997).

\bibitem{tostevin2}
J.~S. Al-Khalili and J.~A. Tostevin, Phys. Rev. C {\bf 57},  1846  (1998).

\bibitem{chulkov}
L.~V. Chulkov, C.~A. Bertulani, and A.~A. Korsheninnikov, Nucl. Phys. A {\bf
  587},  291  (1995).

\bibitem{gareev1}
F.~A. Gareev {\it et~al.}, EuroPhys. Lett. {\bf 20},  487  (1992).

\bibitem{gareev2}
F.~A. Gareev {\it et~al.}, Phys. Atom. Nucl. {\bf 58},  620  (1995).

\bibitem{goncharov}
S.~A. Goncharov and A.~A. Korsheninnikov, Phys. Atom. Nucl. {\bf 58},  1311
  (1995).

\bibitem{gil}
M.~D. Cortina-Gil {\it et~al.}, Nucl. Phys. A {\bf 616},  215c  (1997).

\bibitem{bertulani}
C.~A. Bertulani and H. Sagawa, Nucl. Phys. A {\bf 588},  667  (1995).

\bibitem{korsh1}
A.~A. Korsheninnikov {\it et~al.}, Nucl. Phys. A {\bf 616},  189  (1997).

\bibitem{korsh2}
A.~A. Korsheninnkov {\it et~al.}, Nucl. Phys. A {\bf 617},  45  (1997).

\bibitem{karataglidis}
S. Karataglidis, P.~J. Dortmans, K. Amos, and C. Bennhold, nucl-th {\bf 9811045
  (submitted to Phys. Rev C)},    (1998).

\bibitem{garrido2}
E. Garrido and E.~M. de~Guerra, Nucl. Phys. A {\bf 650},  387  (1999).

\bibitem{garrido3}
E. Garrido, D.~V. Fedorov, and A.~S. Jensen, Nucl. Phys. A {\bf 650},  247
  (1999).

\bibitem{garrido4}
E. Garrido, D.~V. Fedorov, and A.~S. Jensen, Europhys. Lett. {\bf 43},  386
  (1998).

\bibitem{wurzer1}
J. Wurzer and H.~M. Hofmann, Phys. Rev. C {\bf 55},  688  (1997).

\bibitem{hiyama}
E. Hiyama and M. Kamimura, Nucl. Phys. A {\bf 588},  35c  (1995).

\bibitem{varga}
K. Varga, Y. Suzuki, and Y. Ohbayasi, Phys. Rev. C {\bf 50},  189  (1994).

\bibitem{zhukov1}
M.~V. Zhukov {\it et~al.}, Phys. Rep. {\bf 231},  152  (1993).

\bibitem{zhukov2}
M.~V. Zhukov, A.~A. Korsheninnikov, and M.~H. Smedberg, Phys. Rev. C {\bf 50},
  R1  (1994).

\bibitem{corbis}
A. Corbis, D.~V. Fedorov, and A.~S. Jensen, Phys. Rev. Lett. {\bf 79},  2411
  (1997).

\bibitem{kaki}
K. Kaki and S. Hirenzaki, Int. J. Mod. Phys. E {\bf 8},  167  (1999).

\bibitem{density}
C. Elster, S.~P. Weppner, and C.~R. Chinn, Phys. Rev C {\bf 56},  2080  (1997).

\bibitem{thesis}
S.~P. Weppner, Ph.D. thesis, Ohio University, 1997.

\bibitem{sherr}
R. Sherr, Phys. Rev. C {\bf 54},  1177  (1996).

\bibitem{chiral1}
J.~J. Rusnak and R.~J. Furnstahl, Nucl. Phys. A {\bf 627},  495  (1997).

\bibitem{chiral2}
R.~J. Furnstahl, B.~D. Serot, and H.~B. Tang, Nucl. Phys. A {\bf 615},  441
  (1997).

\bibitem{bunny}
B.~C. Clark {\it et~al.}, Phys. Lett. B {\bf 427},  231  (1998).

\bibitem{DH1}
C.~J. Horowitz and B.~D. Serot, Nucl. Phys. A {\bf 368},  503  (1981).

\bibitem{DH2}
C.~J. Horowitz, D.~P. Murdoch, and B.~D. Serot,  in {\em Computational Nuclear
  Physics 1}, edited by K. Langanke, J.~A. Maruhn, and S.~E. Koonin
  (Springer-Verlag, Berlin, 1991).

\bibitem{med2}
C.~R. Chinn, C. Elster, R.~M. Thaler, and S.~P. Weppner, Phys. Rev. C {\bf 52},
   1992  (1995).

\bibitem{med1}
C.~R. Chinn, C. Elster, and R.~M. Thaler, Phys. Rev. C {\bf 48},  2956  (1993).

\bibitem{energydep}
C. Elster and S.~P. Weppner, Phy. Rev. C {\bf 57},  189  (1998).

\bibitem{65mev}
C.~R. Chinn, C. Elster, R.~M. Thaler, and S.~P. Weppner, Phys. Rev. C {\bf 51},
   1418  (1995).

\bibitem{tostevin1}
R. Crespo, J.~A. Tostevin, and R.~C. Johnson, Phys. Rev. C {\bf 51},  3283
  (1995).

\bibitem{garrido1}
E. Garrido, D.~V. Fedorov, and A.~S. Jensen, Nucl. Phys. A {\bf 617},  153
  (1997).

\bibitem{nijmegen}
V.~G. Stokes, R.~A.~M. Klamp, C.~P.~F. Terheggan, and J.~J. de~Swart, Phys.
  Rev. C {\bf 49},  2950  (1994).

\bibitem{cdbonn}
R. Machleidt, F. Sammarruca, and Y. Song, Phys. Rev. C {\bf 53},  R1483
  (1996).

\bibitem{crespo}
R. Crespo and R.~C. Johnson, Phys. Rev. C {\bf 60},  034007  (1999).

\bibitem{thompson}
I.~J. Thompson, J. Phys. G {\bf 23},  1245  (1997).

\bibitem{tanihata}
I. Tanihata {\it et~al.}, Phys. Lett. B {\bf 289},  261  (1992).

\bibitem{zhukov3}
M.~V. Zhukov, A.~A. Korsheninnikov, M.~H. Smedberg, and T. Kobayashi, Nucl.
  Phys. A {\bf 583},  803  (1995).

\bibitem{transfer1}
S.~N. Ershov {\it et~al.}, Phys. Rev. C {\bf 56},  1483  (1997).

\bibitem{coulumb1}
F.~M. Numes and I.~J. Thompson, Phys. Rev. C {\bf 59},  2652  (1999).

\bibitem{coulumb2}
R. Shyam and I.~J. Thompson, Phys. Rev. C {\bf 59},  2645  (1999).

\bibitem{danilin}
B.~V. Danilin {\it et~al.}, Phys. Rev. C {\bf 55},  R577  (1997).

\bibitem{pion1}
S. Karataglidis and C. Bennhold, Phys. Rev. Lett. {\bf 80},  1614  (1998).

\end{thebibliography}

%
%

\end{document}